\documentstyle[12pt]{article}

  \def\pp{{\mathchoice
          {
              \kern 1pt%
              \raise 1pt
              \vbox{\hrule width5pt height0.4pt depth0pt
                    \kern -2pt
                    \hbox{\kern 2.3pt
                          \vrule width0.4pt height6pt depth0pt
                          }
                    \kern -2pt
                    \hrule width5pt height0.4pt depth0pt}%
                    \kern 1pt
           }
            {
              \kern 1pt%
              \raise 1pt
              \vbox{\hrule width4.3pt height0.4pt depth0pt
                    \kern -1.8pt
                    \hbox{\kern 1.95pt
                          \vrule width0.4pt height5.4pt depth0pt
                          }
                    \kern -1.8pt
                    \hrule width4.3pt height0.4pt depth0pt}%
                    \kern 1pt
            }
            {
              \kern 0.5pt%
              \raise 1pt
              \vbox{\hrule width4.0pt height0.3pt depth0pt
                    \kern -1.9pt  
                    \hbox{\kern 1.85pt
                          \vrule width0.3pt height5.7pt depth0pt
                          }
                    \kern -1.9pt
                    \hrule width4.0pt height0.3pt depth0pt}%
                    \kern 0.5pt
            }
            {
              \kern 0.5pt%
              \raise 1pt
              \vbox{\hrule width3.6pt height0.3pt depth0pt
                    \kern -1.5pt
                    \hbox{\kern 1.65pt
                          \vrule width0.3pt height4.5pt depth0pt
                          }
                    \kern -1.5pt
                    \hrule width3.6pt height0.3pt depth0pt}%
                    \kern 0.5pt
            }
        }}

  \def\mm{{\mathchoice
                       {
                             \kern 1pt
               \raise 1pt    \vbox{\hrule width5pt height0.4pt depth0pt
                                  \kern 2pt
                                  \hrule width5pt height0.4pt depth0pt}
                             \kern 1pt}
                       {
                            \kern 1pt
               \raise 1pt \vbox{\hrule width4.3pt height0.4pt depth0pt
                                  \kern 1.8pt
                                  \hrule width4.3pt height0.4pt depth0pt}
                             \kern 1pt}
                       {
                            \kern 0.5pt
               \raise 1pt
                            \vbox{\hrule width4.0pt height0.3pt depth0pt
                                  \kern 1.9pt
                                  \hrule width4.0pt height0.3pt depth0pt}
                            \kern 1pt}
                       {
                           \kern 0.5pt
             \raise 1pt  \vbox{\hrule width3.6pt height0.3pt depth0pt
                                  \kern 1.5pt
                                  \hrule width3.6pt height0.3pt depth0pt}
                           \kern 0.5pt}
                       }}

\catcode`@=11
\def\un#1{\relax\ifmmode\@@underline#1\else
        $\@@underline{\hbox{#1}}$\relax\fi}
\catcode`@=12

\let\du=\du                     

\def\a{\alpha}
\def\b{\beta}

\def\d{\delta}

\def\f{\phi}
\def\g{\gamma}
\def\h{\eta}

\def\l{\lambda}
\def\m{\mu}
\def\n{\nu}

\def\p{\pi}
\def\q{\theta}
\def\r{\rho}

\def\F{\Phi}

\def\L{\Lambda}

\def\ve{\varepsilon}

\def\ck{{\cal K}}

\def\cm{{\cal M}}

\def\co{{\cal O}}

\def\cw{{\cal W}}
\def\cx{{\cal X}}
\def\cy{{\cal Y}}

\def\bo{{\raise-.5ex\hbox{\large$\Box$}}}               
\def\pa{\partial}                                       
\def\pr{\prod}                                          
\def\TH{{\raise.2ex\hbox{$\displaystyle \bigodot$}\mskip-4.7mu \llap H \;}}
\def\face{{\raise.2ex\hbox{$\displaystyle \bigodot$}\mskip-2.2mu \llap {$\ddot
        \smile$}}}                                      

\def\sp#1{{}^{#1}}                              
\def\leftrightarrowfill{$\mathsurround=0pt \mathord\leftarrow \mkern-6mu
        \cleaders\hbox{$\mkern-2mu \mathord- \mkern-2mu$}\hfill
        \mkern-6mu \mathord\rightarrow$}
\def\dvec#1{\vbox{\ialign{##\crcr
        \leftrightarrowfill\crcr\noalign{\kern-1pt\nointerlineskip}
        $\hfil\displaystyle{#1}\hfil$\crcr}}}           
\def\dt#1{{\buildrel {\hbox{\LARGE .}} \over {#1}}}     

\def\frac#1#2{{\textstyle{#1\over\vphantom2\smash{\raise.20ex
        \hbox{$\scriptstyle{#2}$}}}}}                   
\def\sfrac#1#2{{\vphantom1\smash{\lower.5ex\hbox{\small$#1$}}\over
        \vphantom1\smash{\raise.4ex\hbox{\small$#2$}}}} 
\def\bfrac#1#2{{\vphantom1\smash{\lower.5ex\hbox{$#1$}}\over
        \vphantom1\smash{\raise.3ex\hbox{$#2$}}}}       
\def\afrac#1#2{{\vphantom1\smash{\lower.5ex\hbox{$#1$}}\over#2}}    

\def\[{\lfloor{\hskip 0.35pt}\!\!\!\lceil}
\def\]{\rfloor{\hskip 0.35pt}\!\!\!\rceil}
\def\Lag{{\cal L}}
\def\du#1#2{_{#1}{}^{#2}}
\def\ud#1#2{^{#1}{}_{#2}}

\def\fracm#1#2{\hbox{\large{${\frac{{#1}}{{#2}}}$}}}

\def\tr{{\rm tr}}

\def\ul{\underline}
\def\un{\underline}
\def\fracmm#1#2{{{#1}\over{#2}}}

\def\low#1{{\raise -3pt\hbox{${\hskip 0.75pt}\!_{#1}$}}}

\def\Dot#1{\buildrel{_{_{\hskip 0.01in}\bullet}}\over{#1}}
\def\dt#1{\Dot{#1}}

\newskip\humongous \humongous=0pt plus 1000pt minus 1000pt
\def\caja{\mathsurround=0pt}
\def\eqalign#1{\,\vcenter{\openup2\jot \caja
        \ialign{\strut \hfil$\displaystyle{##}$&$
        \displaystyle{{}##}$\hfil\crcr#1\crcr}}\,}
\newif\ifdtup

\def\ref#1{$\sp{#1)}$}

\def\pl#1#2#3{Phys.~Lett.~{\bf {#1}B} (19{#2}) #3}
\def\np#1#2#3{Nucl.~Phys.~{\bf B{#1}} (19{#2}) #3}

\def\pr#1#2#3{Phys.~Rev.~{\bf D{#1}} (19{#2}) #3}

\def\mpl#1#2#3{Mod.~Phys.~Lett.~{\bf A{#1}} (19{#2}) #3}

\parskip=\medskipamount                 
\thicklines                         

\begin{document}


\thispagestyle{empty}               

{\noindent UMDEPP 00--069 \hfill hep-th/0005265 }\\
{\noindent ITP--UH--09/00 \hfill revised version}\\

\noindent
\vskip1.3cm
\begin{center}

{\Large\bf N=1 and N=2 Supersymmetric Non-Abelian}
\vglue.1in
{\Large\bf  Born-Infeld Actions from Superspace~\footnote{Supported in 
part by NSF grant \# PHY--98--02551}}

\vglue.3in

Sergei V. Ketov \footnote{Address after August 1, 2000: Department of Physics,
University of Kaiserslautern, D--67663 \newline ${~~~~~}$ Kaiserslautern, 
Germany}

{\it Department of Physics\\
     University of Maryland\\
     College Park, MD 20742, USA }\\
and \\
{\it Institut f\"ur Theoretische Physik\\
     Universit\"at Hannover, Appelstr.~2\\
     Hannover 30167, Germany}\\
\vglue.1in
{\sl ketov@itp.uni-hannover.de}
\end{center}
\vglue.2in
\begin{center}
{\Large\bf Abstract}
\end{center}

New non-abelian supersymmetric generalizations of the four-dimensional 
Born-Infeld action are constructed in N=1 and N=2 superspace, to all orders in
 $\a'$. The proposed actions are dictated by simple (manifestly supersymmetric
 and gauge-covariant) non-linear constraints.

\newpage

\section{Introduction}

The abelian {\it Born-Infeld} (BI) action is known to be the low-energy part 
of any (gauge-fixed, world-volume) D-brane effective action. The BI action is 
merely dependent upon the abelian vector field strength, being independent 
upon spacetime derivatives of the field strength by definition. The 
supersymmetetric BI actions with linearly realized N=1 or N=2 supersymmetry in
four dimensions describe the low-energy (world-volume) dynamics of a single 
D3-brane propagating in four or six dimensions, respectively. ~\footnote{See 
ref.~\cite{tr} and references therein for a review.} The D3-brane action is 
supposed to have the Goldstone-Maxwell interpretation associated with partial 
(1/2) spontaneous supersymmetry breaking, with the Goldstone fields in a 
(Maxwell) vector supermultiplet with respect to unbroken (linearly realized) 
supersymmetry. Hence, a supersymmetric BI action should be the low-energy 
part of the corresponding Goldstone-Maxwell action. The unbroken N=1 and N=2 
supersymmetry can be made manifest in superspace. The N=1 manifestly 
supersymmetric BI action was first formulated in ref.~\cite{n1}, while its 
Goldstone-Maxwell interpretation was later established in ref.~\cite{n2}. 
The N=2 manifestly supersymmetric BI action was first formulated in 
ref.~\cite{my}, whereas its relation to  (yet to be fully determined) 
N=2 Goldstone-Maxwell action is briefly discussed in sect.~2, 
see also ref.~\cite{my2} for more.

As was pointed out by Witten \cite{wi}, there is a non-abelian
gauge symmetry enhancement when $N$ parallel D3-branes coincide. A 
supersymmetric abelian BI action is then supposed to be replaced by a {\it
Non-abelian Born-Infeld} (NBI) action where the world-volume fields are valued 
in the Lie algebra of $U(N)$.  Both abelian and non-abelian BI actions are 
the {\it effective} actions, defined modulo local field redefinitions. 
Nevertheless, the bosonic abelian BI action (with tension $T_3$)
$$ S_{\rm BI}= -T_3\int d^4x\,\sqrt{-\det\left(\h_{\m\n}+2\p\a' F_{\m\n}\right)
} \eqno(1)$$
is unambiguous, being only dependent of the abelian field strength 
$F_{\m\n}=\pa_{\m}A_{\n} - \pa_{\n}A_{\m}$ but not of spacetime 
derivatives of it $(\pa F)$. In contrast, a bosonic NBI action is not 
well-defined, while there are two principal sources for  ambiguities 
\cite{tr}. The first type of non-abelian ambiguities is related to the obvious
 fact that the terms dependent of 
 the gauge-covariant derivatives of the non-abelian field 
strength cannot be unambiguously separated from the $F$-dependent commutators
since $\[ D_{\m},D_{\n}\]F_{\l\r}=\[F_{\m\n},F_{\l\r}\]$. Any concrete proposal
for an NBI action has to specify an order of the $F$-matrices and, hence, it 
may effectively include some of the $DF$-dependent terms, even if they do not
explicitly appear in the action. Though the full (abelian or non-abelian) 
D3-brane effective action certainly includes the derivative-dependent terms, 
it does not make much sense to keep some of them while ignoring other possible
 terms. Perhaps, the best 
one can do with a bosonic NBI action is to define it for almost covariantly 
constant gauge fields with almost commuting field strengths, which does not
seem to be very illuminating. The second (related) type of ambiguities is 
connected to the trace operation over the gauge group. For example,
 when using the abelian identity $(2\p\a'=b)$
$$ -\det\left(\h_{\m\n}+bF_{\m\n}\right)=1 +\fracmm{b^2}{2}F^2-\fracmm{b^4}{16}
(F\tilde{F})^2~,\quad \tilde{F}^{\m\n}=
\frac{1}{2}\ve^{\m\n\l\r}F_{\l\r}~,\eqno(2)$$
one gets  two natural candidates for the bosonic NBI action, 
$$ S_{\rm (a)}=-T_3\int d^4x\,\sqrt{1+ \fracmm{b^2}{2}\tr(F^2)
-\fracmm{b^4}{16}\tr(F\tilde{F})^2} \eqno(3a)$$
and
$$ S_{\rm (b)}=-T_3\int d^4x\,{\rm Str} \sqrt{-\det(\h_{\m\n}+bF_{\m\n})}~,
\eqno(3b)$$ 
where $F_{\m\n}=F_{\m\n}^at_a$, $\{t_a\}$ are the hermitian generators of the 
gauge group, $\[t_a,t_b\]=if^c_{ab}t_c$, $\tr(t_at_b)=\d_{ab}\,$, and 
${\rm Str}$ is the symmetrized trace,
$$ {\rm Str}\left(t_{a_1}\cdots t_{a_k}\right)=\fracmm{1}{k!}\sum_{\rm 
permutations  \atop \p} \tr\left(t_{\p(a_1)}\cdots t_{\p(a_k)}\right)~.\eqno(4)
$$
The $F$-matrices effectively commute under the symmetrized trace, so that the
formal definition (2) of the determinant still applies in eq.~(3b). It is not
difficult to verify that the equations of motion in the NBI theory (3b) on
self-dual (Euclidean) configurations $(F_{\m\n}=\tilde{F}_{\m\n})$ coincide 
with the ordinary Yang-Mills equations, so that they have {\it the same} BPS
solutions \cite{ha}, though the existence of a BPS bound is not obvious in the
non-abelian case. Away from self-dual configurations the action (3a)  is much
simpler than (3b), while it is also known to admit solitionic (glueball) 
solutions \cite{mos}. 

The gauge-invariant actions (3a) and (3b) are obviously different, so that 
further resolution requirements are needed. Some extra conditions are provided 
by string theory, because the BI action is well-known to represent the 
effective action of slowly varying gauge fields in open string theory. The most
basic requirement of string theory is the overall {\it single} trace of the 
non-abelian gauge field strength products \cite{tr}. The overall 
{\it symmetrized} trace advocated by Tseytlin \cite{tr} is a stronger 
condition based on the observation that it reproduces the $F^4$-terms in the 
non-abelian effective action of open superstrings in ten dimensions \cite{gw}. 
In this Letter I show that adding supersymmetry unexpectedly gives rise to some
more constraints on supersymmetric NBI actions in four dimensions, which are
 not apparent in the bosonic case.

At first sight, it seems to be straightforward to supersymmetrize any NBI 
action, so that supersymmetry would not add anything new towards its intrinsic 
definition. However, in fact, supersymmetry does tell us something more about 
the BI actions. For example, linearly realized supersymmetry apparently 
prefers the parametrization of the abelian BI actions in terms of the 
(anti)self-dual combinations, $F^{\pm}=\frac{1}{2}(F\pm \tilde{F})$, rather 
than in terms of the naively expected tensors $F$ and $\tilde{F}$. More
importantly, it is the spontaneously broken (non-linearly realized) 
supersymmetry on top of the unbroken (linearly realized) supersymmetry 
that is responsible for the complicated non-linear structure of the D3-brane 
action to be considered as the Goldstone-Maxwell action. Hence, the non-linear
structure of the supersymmetric abelian BI actions \cite{tr,n1,my} should also 
 be dictated by similar features. Though a Goldstone interpretation of the 
supersymmetric NBI actions is far from being obvious,  if any,  the 
well-established Goldstone form of the N=1 supersymmetric abelian BI action 
in superspace gives us the natural starting point for a construction of 
its generalizations, either non-abelian or with extended (linearly realized)
 supersymmetry.

The paper is organized as follows. In sect.~2 the N=1 and N=2 supersymmetric
abelian BI actions are reviewed by emphasizing their relation to the 
Goldstone-Maxwell actions. In sect.~3 the non-abelian generalizations of the
BI actions are proposed in N=1 and N=2 superspace. Sect.~4 is my conclusion.
 
\section{N=1 and N=2 abelian BI actions} 

The abelian bosonic BI Lagrangian $\Lag_{\rm BI}(F)$ can be thought of as the 
unique non-linear generalization of the Maxwell Lagrangian, 
$-\fracm{1}{4}F^2$, 
under the conditions of preservation of causality, positivity of energy, and
electric-magnetic duality. In particular, the duality invariance of an abelian
Lagrangian $\Lag(F)$ amounts to the
constraint \cite{sd}
$$ G_{\m\n} \tilde{G}^{\m\n}+F_{\m\n}\tilde{F}^{\m\n}=0~,\quad {\rm where}
\quad \tilde{G}^{\m\n}=\fracm{1}{2}\ve^{\m\n\l\r}G_{\l\r}=
2\fracmm{\pa\Lag}{\pa F_{\m\n}}~.\eqno(5)$$
Supersymmetry is known to be consistent with all these
 physical properties, so 
that the supersymmetric abelian BI actions enjoy similar features.

The N=1 supersymmetric abelian BI action reads \cite{n1}~\footnote{The
deformation parameter $b$ is set to be one. The dependence upon $b$ can
be easily restored \newline ${~~~~~}$ for dimensional reasons. I also ignore
$T_3$ for simplicity.} 
$$ S_{\rm 1BI}=\fracm{1}{2}\left( \int d^4x d^2\q\,W^2 + {\rm h.c.}\right)+\int
d^4x d^4\q\,Y(K,\bar{K})W^2\bar{W}^2~,\eqno(6)$$
where the structure function $Y$ is given by
$$Y(K,\bar{K})=
\fracmm{1}{1-\frac{1}{2}(K+\bar{K})+\sqrt{1-(K+\bar{K})+\frac{1}{4}
(K-\bar{K})^2}}~~~,\eqno(7)$$
and
$$ K=\frac{1}{2}D^2W^2~,\quad D^2=D^{\a}D_{\a}~,\quad W^2=W^{\a}W_{\a}~,\quad
\bar{W}^2=\bar{W}_{\dt{\a}}\bar{W}^{\dt{\a}}~,\eqno(8)$$
in terms of the abelian N=1 chiral spinor superfield strength 
$W_{\a}$, $\a=1,2$, satisfying the off-shell superspace constraints 
(N=1 Bianchi identities)
$$ \bar{D}_{\dt{\a}}W\low{\a}=0~,\quad D^{\a}W_{\a}=
\bar{D}_{\dt{\a}}\bar{W}^{\dt{\a}}~.\eqno(9)$$
The action (6) can be rewritten in 
the `non-linear sigma-model' form 
\cite{n1,tr} 
$$ S_{\rm 1BI}=\int d^4xd^2\q\,\F + {\rm h.c.}~,\eqno(10)$$
where the N=1 chiral Lagrangian $\F$ is the perturbative solution to the
non-linear superfield constraint
$$ \F = \fracm{1}{2}\F\bar{D}^2\bar{\F}+\fracm{1}{2}W^2~.\eqno(11)$$
It is worth mentioning that the constraint (11) is 
Gaussian in $\F$,
while its perturbative solution is unambiguously constructed in superspace 
by iterations.

In fact, the simple constraint (11) is most useful in proving the invariance  
of the action $S_{\rm 1BI}$ under the second (non-linearly realized or
spontaneously broken) supersymmetry with the rigid anticommuting spinor 
parameter $\h^{\a}$ \cite{n1},
$$ \d_2 \F=\h^{\a}W_{\a}~,\quad \d_2W_{\a}=\h_{\a}\left(1-\frac{1}{2}\bar{D}^2
\bar{\F}\right)+i\bar{\h}^{\dt{\a}}\pa_{\a\dt{\a}}\F~,\eqno(12)$$
where the second equation follows from the first one after the use of eqs.~(9)
and (11). The constraint (11) generating the full action (6) is also quite
useful in proving the 
 electric-magnetic self-duality of $S_{\rm 1BI}$. 
The duality invariance amounts to another non-local constraint \cite{kuz}
$$ \int d^4x d^2\q\,(W^2+M^2)=\int d^4xd^2\bar{\q}\,(\bar{W}^2+\bar{M}^2)~,
\quad \fracm{i}{2}M_{\a}=\fracmm{\d S_{\rm 1BI}}{\d W^{\a}}~,\eqno(13)$$
which is the straightforward N=1 generalization of eq.~(5). Thus, 
$S_{\rm 1BI}= S_{\rm GM:~N=2/N=1}$.

Similarly, the N=2 supersymmetric Abelian BI action in N=2 superspace reads
\cite{my}
$$S_{\rm 2BI}=\fracm{1}{2}
\int d^4xd^4\q\,\cw^2+\fracm{1}{4}\int d^4xd^8\q\,\cy(\ck,\bar{\ck})
\cw^2\bar{\cw}^2~,\eqno(14)$$
with {\it the same} structure function
$$ \cy(\ck,\bar{\ck})=\fracmm{1}{1-\frac{1}{2}(\ck+\bar{\ck})
+\sqrt{1-(\ck+\bar{\ck})+\frac{1}{4}(\ck-\bar{\ck})^2}}~,\eqno(15)$$
but
$$ \ck=D^4\cw^2~,\quad D^4=\prod_{i,\a}D^i_{\a}=\fracmm{1}{12}D_{ij}D^{ij}~,
\quad D_{ij}=D^{\a}_iD_{\a j}=D_{ji}~, \quad i=1,2~,\eqno(16)$$
in terms of the N=2 restricted chiral gauge superfield strength $\cw$ 
satisfying the off-shell constraints (N=2 Bianchi identities)
$$ \bar{D}^{\dt{\a}}_i\cw=0~,\quad D^4\cw=\bo\bar{\cw}~.\eqno(17)$$

The action (14) can be rewritten (modulo $\pa W$-dependent terms) in 
the `non-linear sigma-model' form \cite{my2}
$$ S_{\rm 2BI}=\fracm{1}{4}\int d^4x d^4\q\,\cx +{\rm h.c.}~,\eqno(18)$$
whose N=2 chiral Lagrangian $\cx$ satisfies the non-linear 
N=2 superfield constraint
$$ \cx=\fracm{1}{4}\cx\bar{D}^4\bar{\cx}+\cw^2~.\eqno(19)$$

Similarly to the N=1 abelian BI action, the non-linear constraint (19) gives 
us the convenient way of handling the complicated N=2 BI abelian 
action (14). For example, as was demonstrated in ref.~\cite{kuz}, 
electric-magnetic self-duality of an N=2 action $S(\cw,\bar{\cw})$  amounts 
to the following N=2 supersymmetric extension of the N=1 non-local constraint 
(13):
$$  \int d^4xd^4\q\,(\cw^2+\cm^2)=\int d^4x d^4\bar{\q}\,(\bar{\cw}^2+
\bar{\cm}^2)~,\quad \fracm{i}{4}\cm=\fracmm{\d S}{\d \cw}~,\eqno(20)$$
while it appears to be satisfied in the case of $S_{\rm 2BI}$ defined by
eqs.~(18) and (19).
 
Unlike its N=1 BI counterpart, the N=2 BI action (14) or (18) does not give 
the full N=2 Goldstone-Maxwell action, but rather represents its low-energy 
part, i.e. $S_{GM:~N=4/N=2}=S_{\rm 2BI}+\co(\pa W,\pa\bar{W})$.
The infinitesimal parameters of the spontaneously broken (non-linearly 
realised) rigid symmetries can be naturally unified into a single 
(spacetime-independent) superfield
$\L=\l + \q^{\a}_i\l^i_{\a} + \q_{ij}\l^{ij}$, where $\l$ is the complex 
parameter of the Peccei-Quinn-type symmetry 
associated with two spontaneously broken translations (from the the viewpoint
 of a D3-brane propagating in six dimensions), $\l^i_{\a}$ are the spinor
parameters of two spontaneously broken supersymmetries, whereas $\l^{ij}$ are
the parameters of spontaneously broken R-symmetry $SU(2)$. The natural 
{\it Ansatz} for the transformation laws of the non-linearly realised 
symmetries is given by an N=2 analogue of eq.(12) as follows \cite{my2}:
$$ \d_2\cx=2\L\cw~,\quad  
\d_2\cw=\L\left(1-\fracm{1}{4}\bar{D}^4\bar{\cx}\right)+\ldots,\eqno(21)$$
where $\cx$ is the perturbative solution to the non-linear constraint (19) by
iterations, to all orders in $\cw$ and $\bar{\cw}$, while the dots stand for
some $\pa W$-dependent terms needed for consistency with the second Bianchi 
identity (17). A variation of the N=2 BI action (18) under the non-linear 
transformations (21) does not vanish, but it appears to be only dependent 
upon higher spacetime derivatives of $W$ and $\bar{W}$. This may not be 
surprising since the BI action and its supersymmetric extensions were defined 
modulo such terms. However, it also means that the N=2 BI action has to be 
modified, order by order in $\pa W$ and $\pa\bar{W}$, in order to get the 
full N=2 Goldstone-Maxwell action. A derivation
 of the derivative-dependent terms is beyond the scope of this Letter, and we 
do not need them for our main purpose formulated in the title.
 
A manifestly N=4 supersymmetric abelian BI action is not known (see, however,
ref.~\cite{tr} and references therein for some partial results).

\section{N=1 and N=2 supersymmetric NBI actions}

Having understood the fact that the simple non-linear constraints (11) and (19)
fully determine the structure of the highly complicated abelian BI actions (6) 
and (14), respectively,   it is natural to {\it define} the N=1 and N=2
supersymmetric NBI actions by non-abelian generalizations of eq.~(11) and (19).

The non-abelian (Yang-Mills) N=1 chiral superfield strength is given by the 
well-known formula (see, e.g., ref.~\cite{sohn} for a review or 
an introduction)~\footnote{All superfields are now Lie algebra-valued.}
$$ W_{\a}=\fracm{1}{8}\bar{D}^2\left( e^{-2V}D_{\a}e^{2V}\right)~,\quad
\bar{D}_{\dt{\a}}W\low{\a}=0~,\eqno(22)$$ 
where the real scalar gauge superfield potential $V$ transforms under gauge 
transformations with the chiral paramater $\L(x,\q,\bar{\q})$ in the
standard way \cite{sohn}:
$$ e^{2V}\to e^{-2i\bar{\L}}e^{2V}e^{2i\L}~,\quad \bar{D}_{\dt{\a}}\L=0~,
\eqno(23)$$
so that $W_{\a}$ and $\bar{W}_{\dt{\a}}$ transform covariantly, {\it viz.}
$$ W_{\a}\to e^{-2i\L}W_{\a}e^{2i\L}~,\quad \bar{W}_{\dt{\a}}\to
e^{-2i\bar{\L}}\bar{W}_{\dt{\a}}e^{2i\bar{\L}}~~~.\eqno(24)$$

The non-abelian gauge-covariant generalization of eq.~(11) is given by
$$ \F=\fracm{1}{2}\F\bar{D}^2\left(e^{-2V}\bar{\F}e^{2V}\right)
+\fracm{1}{2}W^2~,\eqno(25)$$
where $\F$ is the N=1 chiral superfield Lagrangian that transforms like 
$W_{\a}$ under the gauge transformations. The invariant action reads
$$S_{\rm 1NBI}= \int d^4x d^2\q\,\tr\, \F + {\rm h.c.}~,\eqno(26a)$$
or
$$ S^{(s)}_{\rm 1NBI} =\int d^4xd^2\q\,{\rm Str}\, \F +{\rm h.c.} \eqno(26b)$$
The NBI actions (26a) and (26b) are supersymmetric and gauge-invariant, while  
they both have the 
 single overall trace. The symmetrized trace in eq.~(26b) is 
supposed to be applied to the gauge-covariant operators only, by definition.

It is instructive to take a look at the structure of the quartic $(F^4)$ terms 
in the actions (26), which arise from the 
standard `adjoint chiral matter' term,
$$ \int d^4x d^4\q \,{\rm (S)tr}\, \f e^{-2V}\bar{\f}e^{2V}~,\quad \f=W^2~.
\eqno(27)$$ 
It is straightforward to verify that taking the trace as in eq.~(26a) 
  results in the non-abelian generalization of the Euler-Heisenberg 
 Lagrangian, in the bosonic sector of the component expansion of the action 
(26a),
$$ \fracm{1}{4}\left[ \tr(F^2)^2  +\tr(F\tilde{F})^2\right]~.\eqno(28)$$
In contrast, taking the symmetrized trace, as in eq.~(26b), exactly yields the 
$F^4$-terms appearing in the expansion of the bosonic NBI Lagrangian (3b)
\cite{arg}. Hence, if one insists on the choice (3b) of the bosonic NBI 
action, its supersymmetric extension in 
compact form is provided by 
eq.~(26b) --- {\it cf.} ref.~\cite{arg}. Supersymmetry alone does not provide 
a resolution between the two different actions (26a) and (26b), so that more
physical input is apparently needed. One natural option is to restrict the
gauge group $G$ to its (abelian) Cartan subgroup and then impose the condition
of the generalized (non-abelian) self-duality for the resulting supersymmetric
 field theory of rank $G$ (abelian) gauge fields along the lines of 
ref.~\cite{zu}. As was argued in ref.~\cite{zu}, this may ultimately support 
the symmetrized trace. Still, in the absence of more physical reasons, 
 the ordinary trace in eq.~(26a) is much simpler, being dependent of  only two
 matrix building blocks, $W^2$ and $\bar{W}^2$ (or $F^2$ and $F\tilde{F}$), 
and their covariant derivatives (see below). The action (3a) does not seem to 
have a nice supersymmetric generalization.

It is possible to rewrite the action (26) into the manifestly 
gauge-invariant and N=1 supersymmetric form, by using the N=1 supersymmetric 
gauge-covariant derivatives in superspace, which satisfy the standard N=1 
super-Yang-Mills constraints \cite{sohn}
$$ \{ \nabla\low{\a}, \nabla\low{\b} \} = \{ \bar{\nabla}_{\dt{\a}},
\bar{\nabla}_{\dt{\b}} \} =0~,\quad 
\{ \nabla\low{\a}, \bar{\nabla}_{\dt{\b}} \}
=-2i\nabla_{\a\dt{\b}}~,$$
$$ \[ \nabla\low{\a},\nabla_{\b\dt{\b}} \] = 
2i\ve\low{\a\b}\hat{\bar{W}}_{\dt{\b}}~,
\quad
\[ \bar{\nabla}_{\dt{\a}}, \nabla_{\b\dt{\b}} \] =2i\ve_{\dt{\a}\dt{\b}}
\hat{W}\low{\b}~,\eqno(29)$$
where $\hat{W}_{\a}$ is the N=1 covariantly-chiral gauge superfield strength,
$$ \bar{\nabla}_{\dt{\a}}\hat{W}\low{\a}=0~,\quad \bar{\nabla}_{\dt{\a}}
\hat{\bar{W}}{}^{\dt{\a}}=\nabla^{\a}\hat{W}_{\a}~.\eqno(30)$$  
Equation (26) then takes the form
$$ S_{\rm 1NBI}= \int d^4xd^2\q\, {\rm (S)tr}\,\hat{\F} + {\rm h.c.}~,
\eqno(31)$$
where the N=1 covariantly-chiral Lagrangian $\hat{\F}$ is the perturbative
(iterative) solution to the manifestly gauge-covariant and supersymmetric 
nonlinear superfield constraint
$$ \hat{\F}=\fracm{1}{2} \hat{\F}\bar{\nabla}^2\hat{\bar{\F}}
+\fracm{1}{2}\hat{W}^2 ~.\eqno(32)$$

It is not difficult to generalize eqs.~(31) and (32) further to the case of N=2
supersymmetry, by doing a similar construction in N=2 superspace. The standard
N=2 superspace constraints, defining the off-shell N=2 supersymmetric 
Yang-Mills theory, are given by \cite{gsw}
$$\eqalign{
 \{  \bar{\nabla}_{\dt{\a}i}, \bar{\nabla}_{\dt{\b}j} \} =
-2\ve_{\dt{\a}\dt{\b}}\ve\low{ij}\hat{\cw}~,\quad & \quad
\{ \nabla^i_{\a},\nabla^j_{\b} \} = 
-2\ve\low{\a\b}\ve^{ij}\hat{\bar{\cw}}~,\cr 
\[ \nabla_{\a\dt{\a}},\bar{\nabla}_{\dt{\b}i} \] = i\ve_{\dt{\a}\dt{\b}}
\nabla\low{\a i}\hat{\cw}~,\quad & \quad
\[ \nabla_{\a\dt{\a}},\nabla^i\low{\b} \] =
i\ve\low{\a\b}\bar{\nabla}^i_{\dt{\a}}\hat{\bar{\cw}}~,\cr}$$
$$ \{ {\nabla^i}\low{\a}, \bar{\nabla}_{\dt{\b}j} \} =-2i\d\ud{i}{j}
\nabla_{\a\dt{\b}}~~,\eqno(33)$$
where the non-abelian N=2 gauge superfield strength $\hat{\cw}$ obeys the
off-shell constraints (N=2 Bianchi identities)
$$ \bar{\nabla}_{\dt{\a}i}\hat{\cw}=0\quad{\rm and}\quad 
\nabla_{ij}\hat{\cw}=\bar{\nabla}_{ij}\hat{\bar{\cw}}~~.\eqno(34)$$
I use the following book-keeping notation:
$$ \eqalign{
\nabla_{ij}=\nabla^{\a}_{(i}\nabla_{j)\a}=\nabla_{ji}~,\quad & \quad
\bar{\nabla}\low{ij}=\bar{\nabla}_{\dt{\a}(i}\bar{\nabla}^{\dt{\a}}\low{j)}=
\bar{\nabla}\low{ji}~,\cr
\nabla_{\a\b}=\nabla_{i(\a}\nabla^i_{\b)}=\nabla_{\b\a}~,\quad & \quad
\bar{\nabla}_{\dt{\a}\dt{\b}}=
\bar{\nabla}\low{i(\dt{\a}}\bar{\nabla}^i_{\dt{\b})}=
\bar{\nabla}_{\dt{\b}\dt{\a}}~,\cr}\eqno(35)$$
where all symmetrizations have unit weight. The non-abelian generalization 
of the $\bar{D}^4$ operator, which converts the (covariantly) anti-chiral N=2
superfields into (covariantly) chiral N=2 superfields, is most easily (and
unambiguously) identified in the $SL(4,{\bf C})$ notation of ref.~\cite{sg},
by combining 
fundamental $SL(2,{\bf C})$ and $SU(2)$ indices into a single
(fundamental) $SL(4,{\bf C})$ index $\ul{a}=(\dt{\a},i)=1,2,3,4$. The
$\bar{\nabla}$-algebra of eq.~(33) in the $SL(4,{\bf C})$ notation takes the 
familiar Dirac-type-form
$$ \{ \bar{\nabla}_{\ul{a}},\bar{\nabla}_{\ul{b}} \} =2C_{\ul{a}\ul{b}}
\hat{\cw}~,\qquad  \bar{\nabla}_{\ul{a}}\hat{\cw}=0~,\eqno(36)$$
with the constant metric $C$, $C^2=1$ and $C^T=C$.  The desired gauge-covariant
operator  is just given by the `$\g_5$-type' top product
$$ \bar{\nabla}^4= \fracmm{1}{4!} \ve^{\ul{a}\ul{b}\ul{c}\ul{d}}
\bar{\nabla}_{\ul{a}}\bar{\nabla}_{\ul{b}}\bar{\nabla}_{\ul{c}}
\bar{\nabla}_{\ul{d}}~~.\eqno(37)$$
In the notation (35) it reads
$$ \bar{\nabla}^4= \fracmm{1}{24}\left( \bar{\nabla}\low{ij}\bar{\nabla}^{ij}
-\bar{\nabla}_{\dt{\a}\dt{\b}}\bar{\nabla}^{\dt{\a}\dt{\b}}\right)
-\fracmm{2}{3}\hat{\cw}^2~~.\eqno(38)$$

The N=2 supersymmetric non-abelian Born-Infeld action is given by
$$ S_{\rm 2NBI}= \fracm{1}{4} \int d^4xd^4\q\,{\rm (S)tr}\,\hat{\cx}
+{\rm h.c.}~~,\eqno(39)$$
whose N=2 covariantly chiral Lagrangian $\hat{\cx}$ is the perturbative
(iterative) solution to the N=2 superfield constraint
$$ \hat{\cx} = \fracm{1}{4}\hat{\cx}\bar{\nabla}^4 \hat{\bar{\cx}} + 
\hat{\cw}^2~.\eqno(40)$$

\section{Conclusion}

The proposed N=1 and N=2 supersymmetric NBI actions in components contain only
even powers of $F$, while they reduce to the known super-Born-Infeld  actions 
in the abelian case. Both actions enjoy `auxiliary freedom' by keeping the 
auxiliary fields $\vec{D}$ (in the  Wess-Zumino  gauge) away from propagation, 
 with $\vec{D}=0$ being a  solution to their equations of motion. 

Unlike the supersymmetric abelian BI actions (sect.~2), their supersymmetric
non-abelian counterparts (sect.~3) are dependent  of the gauge superfields
not only via their gauge superfield strengths but also directly (via the 
gauge-covariant derivatives). This does not allow us to extend the notion of 
abelian electric-magnetic duality to the supersymmetric NBI actions. 
Similarly, it is unclear to us whether our supersymmetric non-abelian BI 
actions admit any Goldstone-Yang-Mills interpretation.

It would be also interesting to investigate the  structure of BPS solutions 
to the new supersymmetric NBI actions and find a precise relation between 
these actions and the non-abelian Dirac-Born-Infeld actions describing 
clusters of D3-branes with `deformed' (non-linear) supersymmetry. 
A connection to noncommutative geometry seems to exist along the lines of 
ref.~\cite{sw} too. 

\section*{Acknowledgement} I thank S. J. Gates Jr. and E. A. Ivanov for 
discussions. I also acknowledge kind hospitality extended to me at the 
University of Maryland in College Park during preparation of this paper.

\end{document}